# A Novel Strategy to Strengthen Directionally Solidified Superalloy Through Grain Boundary Simplified Design

Yunpeng Fan[a], Xinbao Zhao[a,b,*], Yu Zhou[a], Quanzhao Yue[a], Wanshun Xia[a,*], Yuefeng Gu[a,b,*], Ze Zhang[a,b]

[a] Institute of Superalloys Science and Technology, School of Materials Science and Engineering, Zhejiang University, Hangzhou 310027, China

[b] State Key Laboratory of Silicon and Advanced Semiconductor Materials, School of Materials Science and Engineering, Zhejiang University, Hangzhou, 310027, China

* Corresponding authors.

E-mail addresses: superalloys@zju.edu.cn (X.B. Zhao), guyf@zju.edu.cn (Y.F. Gu), wanshunxia@zju.edu.cn (W.S Xia)

## Abstract

Conventional strategies for enhancing creep resistance often rely on grain boundary strengthening, yet this approach can inadvertently promote premature grain boundary fracture. This study presents a subtractive alloy design strategy for nickel-based directionally solidified superalloys (DS superalloy) through elimination of conventional grain boundary strengthening elements (C, B, Zr) and the strategy improves the creep performance by 60% rivaling 2$^{nd}$ generation single crystal superalloys. Through characterization of heat-treated and heat-exposed microstructures, we confirm the suppression of deleterious grain boundary phases. Creep tests and fracture analysis reveal a critical transition in failure mechanism: the removal of these elements shifts the fracture mode from transgranular to intergranular. Our discussion comprehensively links this microstructural engineering to the underlying creep deformation mechanisms, showing that the enhanced performance stems from stabilization of γ channel and phase interfaces within grains, as well as strengthening of grain boundaries through serration. This work establishes a novel materials design principle that decouples grain boundary strengthening from elemental additions, offering transformative potential for next-generation high-efficiency turbine blade applications.

# 1. Introduction

Nickel-based directionally solidified superalloys (DS superalloy) are widely employed as turbine blade materials for land-based gas turbines due to their superior performance compared to conventional cast superalloys and lower manufacturing costs relative to single crystal superalloy (SC superalloy)[1–5]. The strengthening mechanisms of these alloys generally involve two aspects: intragranular strengthening and grain boundary strengthening. Generally, intragranular strengthening is achieved through solid solution strengthening and precipitation strengthening[6], whereas grain boundary strengthening primarily relies on the addition of boundary strengthening elements such as C, B, and Zr. These elements facilitate the formation of precipitated phases at grain boundaries, effectively impeding grain boundary migration[7,8]. Although they profoundly affect the formation and evolution of the intragranular/grain boundary microstructure and may bring long-term stability risks while strengthening the grain boundaries, addition of grain boundary strengthening elements remains indispensable in conventional alloy design owing to that creep damage originates from sliding[9] and cavitation[10,11] of grain boundaries (GB) in high temperature/stress coupling environment.

In most mainstream nickel-based DS superalloys, grain boundary strengthening is universally achieved through deliberate incorporation of boundary strengthening elements, resulting in carbon-containing carbide phases serving as the primary precipitates at grain boundaries[12–14]. Furthermore, boron and zirconium elements have been observed to influence both the morphology and dimensional characteristics of these carbide phases[15–19]. As for the grain boundary strengthening, variations in carbon content primarily affect three aspects: phase composition, dendritic segregation, and thermophysical properties of the alloy[20–22]. Subsequently, these changes influence the quantity and dimensions of carbides within the material[23]. For grain boundary strengthening, increased carbide formation can inhibit grain boundary sliding and deformation while promoting dislocation network formation at phase interfaces during high-temperature and low-stress creep[20,22], it may simultaneously serve as potential crack initiation sites[23]. However, excessive carbon content tends to deplete Ta in interdendritic regions, leading to precipitation of brittle topologically close-packed (TCP) phases that induce premature stress concentration and fracture. Moreover, Li et al.[24]further identified that carbide formation might reduce matrix strength through consumption of solid solution strengthening elements in the alloy. Consequently, the influence of grain boundary strengthening elements on grain boundaries and intragranular structures should be comprehensively considered, and the strength of grain boundaries and intragranular structures should be synergistically regulated.

This study proposes a "Grain Boundary Simplification" design strategy, innovatively introducing the concept of eliminating grain boundary strengthening elements in directionally solidified superalloys to enhance rupture life specifically under high-temperature and low-stress conditions. The investigation revealed that although the removal of grain boundary strengthening elements intensified dendritic segregation, it significantly increased the volume fraction of intragranular γ′ precipitates, enhanced γ/γ′ lattice misfit and improved microstructural stability. Crucially, this approach not only elevates intragranular strength but also enhances grain boundary strength. This dual enhancement achieved better mechanical compatibility between grain interiors and boundaries, collectively enhancing the alloy's stress rupture resistance under high-temperature and low-stress service conditions. This design strategy enables low-cost Re-free directionally solidified alloys to surpass the creep performance of Re-containing second-generation single-crystal superalloys.

## 2. Experimental procedures

The compositions of two independently designed DS superalloys were presented in Table 1. The DS-GBFree alloy represents a directionally solidified variant derived from the DS-Base alloy by eliminating grain boundary strengthening elements including C, B, and Zr, with their concentrations determined with reference to [25–28]. The DS rods were solidified via the Bridgeman method. After solidification, the DS rods were fully heat treated via a solution treatment and two aging treatments at relatively low temperature, and the temperature in solution treatment was determined by the method of incipient melting and differential scanning calorimetry (DSC). Based on phase diagram calculations and incipient melting experiments, both alloy samples were heated to 1300°C directly, and were kept for hours to achieve sufficient elemental homogenization during the solid solution heat treatment. At the stage of solid solution treatments, the samples were held in the Ar atmosphere to prevent oxidation at extremely high temperature. The heat exposure experiments were applied to the fully heat-treated specimens, and the method used to calculate the size of the γ′ phases after heat treatment was consistent with that reported in[29].

**Table 1** Nominal compositions of the Ni-DS investigated in this work (wt.%).

| Element | Cr + Co | Al + Ti | Mo | W | Ta | Nb | Hf | C | B | Zr | Ni |
|---|---|---|---|---|---|---|---|---|---|---|---|
| DS-Base | 12.0 | 6.0 | 3.0 | 12.0 | 4.0 | 0.5 | 0.1 | 0.1 | 0.01 | 0.01 | Bal. |
| DS-GBFree | 12.0 | 6.0 | 3.0 | 12.0 | 4.0 | 0.5 | 0.1 | 0 | 0 | 0 | Bal. |

After full heat treatments, the samples were machined to the dumbbell shape with a gauge length of 25 mm and a diameter of 5 mm to carry out the creep tests at 1040°C and 137MPa, which closely approximated the service environment of first-stage blades in land-based gas turbines, where temperature capability at 137 MPa conventionally served as the criterion for generational classification of superalloys. The temperature fluctuation range during the creep tests was controlled within 3°C. After rupture, the samples showed typical necking regions. To investigate the microstructural evolution in detail, we took samples at different sites from the fracture surface. Regions with 1 mm, 5 mm, 10 mm and 20 mm distance to the fracture surface were studied to compare the effects of different strain levels. The cross-section samples at different distances from the fracture T1(2mm) and T2(10mm) were sliced to prepare samples of each condition for transmission electron microscope (TEM) observation of microscopic defects. The longitudinal-section samples at different distances from the fracture S1, S2 and S3 were selected for scanning electron microscope (SEM) observation to study the microstructure evolution after creep rupture.

Specimens for SEM observation were etched by a solution consisting of HF, $HNO_3$ and glycerin (volume ratio 1:2:3) after mechanically grounded and polished. The SEM images were acquired from FEI Quanta 650. The measurement methodology for dendritic segregation coefficients aligned with established protocols in[30]. For both alloys, the as-cast segregation coefficients were determined through localized sampling adjacent to eutectic structures in dendritic cores and interdendritic regions. Post-heat-treatment segregation analysis employed a systematic 10×10 grid point method, ensuring comprehensive spatial coverage across the microstructure. This dual-scale characterization approach enables quantitative evaluation of elemental redistribution during thermal processing. The specimens for Electron Backscatter Diffraction (EBSD) was polished in colloidal silica for 4h. The specimens for transmission electron microscope (TEM) observation were prepared by the electrochemical double jet thinning method at a voltage of about 15V and temperature of about -30°C. The TEM observation was obtained by FEI Tecnai G2 F20 S-TWIN. The thermodynamic calculation was carried out on Thermo-Calc, using the database of TCNi9: Ni-Alloys v9.1.

To quantify the grain boundary serration at different distances from the fracture surface, a series of SEM images were captured at a magnification of 2500x along the grain boundaries. These images were then stitched together to obtain a complete profile of the grain boundary. The position of the grain boundary was mapped and converted into a function relative to the distance from the fracture surface. After that, the grain boundary position function was detrended to remove any linear components. At last, the function was subjected to a Fast Fourier Transform (FFT) to derive the frequency-amplitude relationship. Further analysis was conducted to generate a wavelength-accumulated-amplitude relationship graph. In the results of the FFT, the wavelength quantified the degree of curvature of the grain boundary, while the amplitude represented the prevalence of that particular wavelength component within the grain boundary. A high amplitude at shorter wavelengths indicated that the wavelengths of the serrated grain boundary were larger, and conversely, higher amplitudes at longer wavelengths indicated smaller wavelengths of the serrated grain boundary.

## 3. Results

*3.1 Initial microstructure before and after heat treatments*

Figure 1 illustrated the microstructure of DS-GBFree and DS-Base alloys in both as-cast and heat-treated conditions. There were obvious cross dendrites in both cast alloys, as shown by the white dashed lines in Figure 1(a) and (g). In the as-cast condition, both alloys exhibited a cruciform morphology of γ′ phase within the dendrite cores, as shown in Figure 1(b) and (h). The characteristic eutectic structures formed in interdendritic regions due to segregation of low-melting elements like Al and Ti during solidification were indicated by yellow arrows in Figure 1(a)(g) and the enlarged images were shown in the Figure 1(c)(i). Quantitative analysis revealed distinct eutectic area fractions of 1.4% for DS-GBFree alloy versus 0.9% for DS-Base alloy, demonstrating that the incorporation of grain boundary strengthening elements effectively suppresses eutectic formation[31]. This suppression mechanism related to the development of skeleton-like MC carbides (red arrows and orange box in Figure 1(d)), which will be further discussed in subsequent sections.

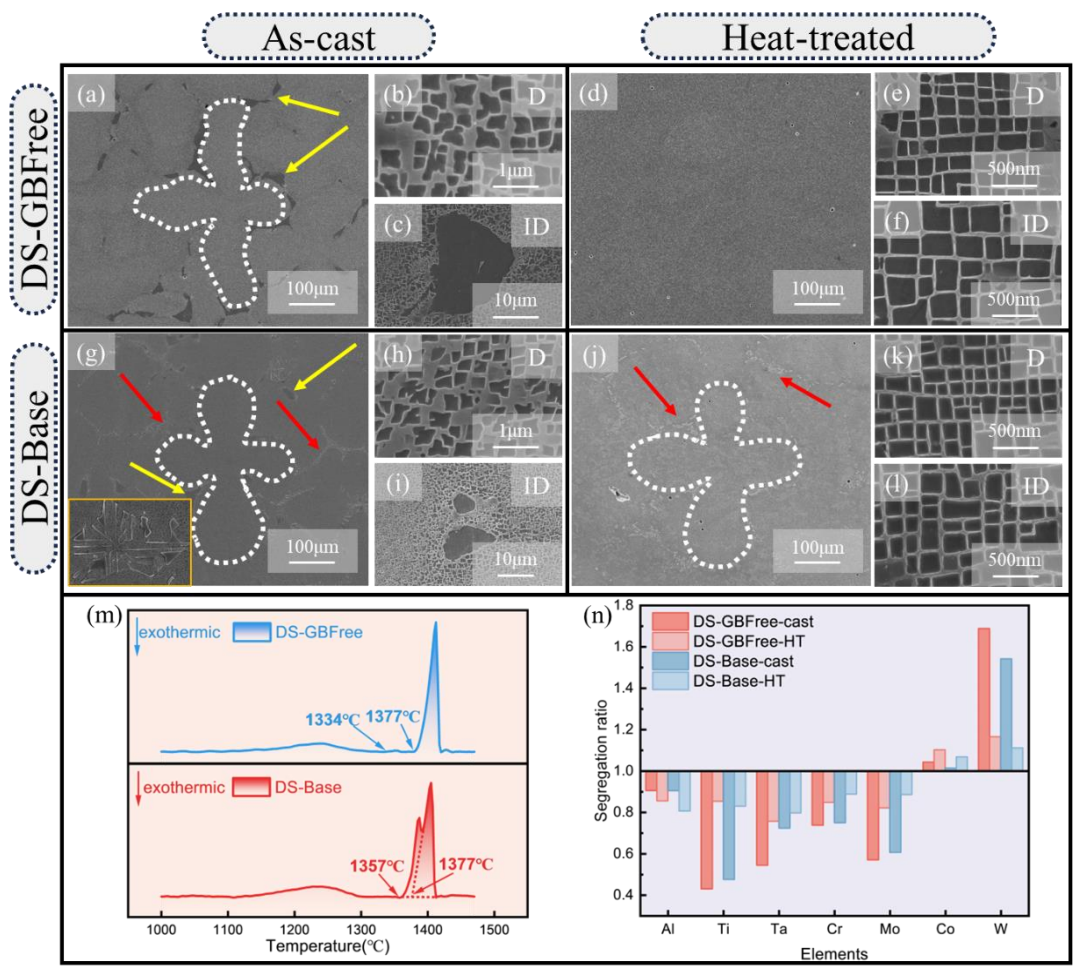

**Figure 1** The microstructure of (a-c) as-cast and (d-f) heat-treated DS-GBFree alloy in dendritic cores and interdendritic regions. (d) The microstructure of (g-i) as-cast and (j-l) heat-treated DS-Base alloy in dendritic cores and interdendritic regions. The (m) Heating DSC curves and (n) dendritic segregation coefficients under as-cast and heat-treated conditions of DS-GBFree and DS-Base. (The DSC data of DS-GBFree was adapted and re-plotted from Ref.[32]).

The DSC curves of both alloys were presented in Figure 1(m). Due to the presence of GB strengthening elements in the DS-Base alloy, it exhibited two prominent endothermic peaks at 1357°C (MC carbide dissolution) and 1377°C (γ phase dissolution), while DS-GBFree alloy showed a single dominant γ phase peak at 1377°C with an additional eutectic melting peak at 1334°C. Consequently, the solution treatment for each alloy was optimized to 1300°C, and the aging treatments were set as 1100°C/4h + 870°C/16h. As shown in Figure 1(d) and (j), the eutectic structures in both alloys were completely eliminated after full heat treatment, while the γ′ phases in dendrite cores and interdendritic regions developed regular cuboidal morphologies as shown in Figure 1(e)(f)(k)(l). However, carbide phases persisted in the DS-Base alloy, indicated by red arrows in Figure 1(j) due to their dissolution temperature measured by DSC being higher than the solution treatment temperature. These retained carbides outline the dendritic morphology in the heat-treated DS-Base alloy, as demarcated by white dashed lines in Figure 1(j).

The influence of carbon on dendritic segregation remained controversial in previous research[24,33]. In this study, both alloys exhibited comparable two-phase microstructures and segregation coefficients after heat treatment. It is noteworthy that the DS-GBFree alloy exhibited a slightly higher segregation coefficient than the DS-Base alloy, which was associated with carbides and would be discussed in subsequent sections. The observed increase in segregation coefficients for Al and Co after heat treatment arises from differences in composition analysis: cast state measurements employed selective point analysis that deliberately excluded eutectic constituents, with interdendritic sampling positioned adjacent to eutectic regions, thereby underestimating Al content. For Co element, solidification-induced rejection during eutectic formation enriched surrounding regions, creating composition profiles resembling dendrite cores and consequently depressing measured segregation coefficients[32].

A detailed comparative analysis of γ/γ′ microstructures in heat-treated DS-GBFree and DS-Base alloys was conducted to evaluate the effects of grain boundary strengthening element elimination. The intragranular area fractions of γ′ phases in two alloys were 75.0±0.5% and 69.0±0.6%, respectively. Figure 2 presented the statistically quantified γ′ precipitate size distributions in dendritic cores. Gaussian function fitting was applied to all histograms to characterize the distribution patterns systematically. Although the γ' phase dimensions in interdendritic regions differed significantly between the two alloys (Figure 1(f) and (l)), their dendrite cores exhibited comparable average sizes and full width at half maximum (FWHM) values, as shown in Figure 2(c) and (d). However, substantial differences existed in γ channel widths within dendrite cores. The DS-GBFree alloy displays wider γ channels, which reduced resistance to dislocation motion in the γ phase and was generally detrimental to creep performance[34].

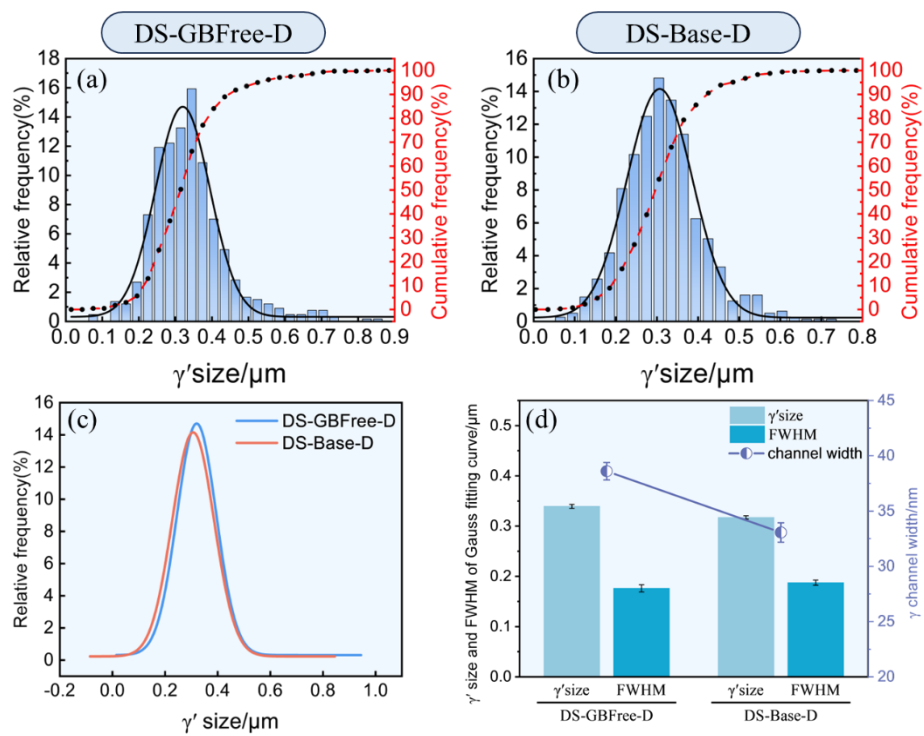

**Figure 2** Statistical histograms of γ′ precipitate in dendritic core of (a) DS-GBFree and (b) DS-Base. (c) Gaussian fitting curves of the size distributions in the dendritic core across both alloys. (d) Average γ′ phase dimensions (light blue) and full width at half maximum (FWHM, dark blue) of Gaussian distributions and γ phase channel widths of both alloys.

Figure 3 revealed the critical role of MC carbides in microstructural evolution in DS-Base alloy. In as-cast DS-Base alloy, Ta/Ti/Nb-rich carbides exhibited distinct morphologies, skeleton-like formations in intragranular region (Figure 3 (a-d)) versus blocky configurations in grain boundary (Figure 3 (e-h)). High-resolution transmission electron microscopy (HRTEM) characterization in Figure 3 (i) and selected area electron diffraction (SAED) in Figure 3 (j) confirmed the face-centered cubic (FCC) structure of these carbides, with measured lattice spacing of 0.226 nm for {020} planes corresponding to a unit cell parameter a = 0.452 nm. The thermal stability of these carbides proved remarkable, collected Ti and Ta to prevent the diffusion and movement, persisting through heat treatment as shown in Figure1(j) to create fundamental differences in segregation behavior between the two alloy. The effect of the carbides on the mechanical stability and mechanical properties of the alloy would be discussed later.

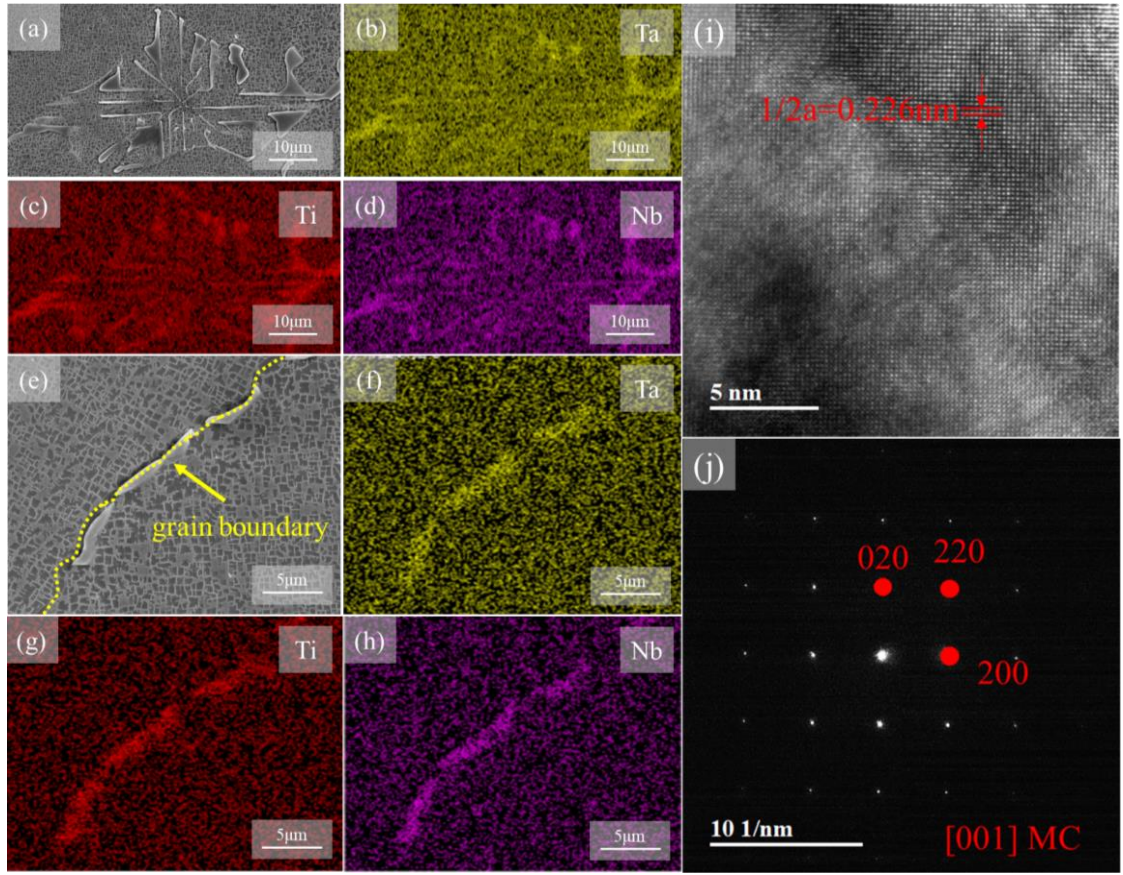

**Figure 3** (a-d) The morphology and energy dispersive spectrometer (EDS) mapping of skeletal-like MC carbide in intragranular region. (e-h) The morphology and EDS mapping of blocky MC carbide in grain boundary. (i) HRTEM image of MC carbide in DS-Base alloy. (j) SAED image of MC carbide in DS-Base alloy.

*3.2 Microstructural evolution during thermal exposure*

To investigate the influence of grain boundary strengthening element removal on microstructural stability, both alloys underwent thermal exposure at 1040°C for 100 h and 500 h. As demonstrated in Figure S1, morphological disparities in γ′ precipitates between dendrite cores and interdendritic regions during thermal exposure were smaller than variations induced by either exposure duration or alloy composition. Consequently, only dendrite core morphologies were selected for detailed quantitative analysis. Figure 4 illustrated the γ/γ′ microstructures in dendritic cores and corresponding γ′ precipitate size distributions with Gaussian fitting curves for all four conditions, while Figure 4 (i-

k) summarize the fitting parameters including average γ′ size, full width at half maximum (FWHM), and γ-phase channel width. Both alloys exhibited significant coarsening compared to their heat-treated states, with more pronounced effects after 500 h exposure versus 100 h. Notably, Gaussian fitting was employed to analyze the precipitate size distribution (PSD) results, with the full width at half maximum (FWHM) quantified for all specimens. Enhanced FWHM values indicated broader γ′ size distribution peaks, reflecting increased microstructural heterogeneity. The elimination of grain boundary strengthening elements was found to promote FWHM expansion in γ′ distributions - a critical observation given that homogeneous γ′ precipitates generally enhance mechanical properties, while heterogeneous distributions may induce stress concentration risks[35].

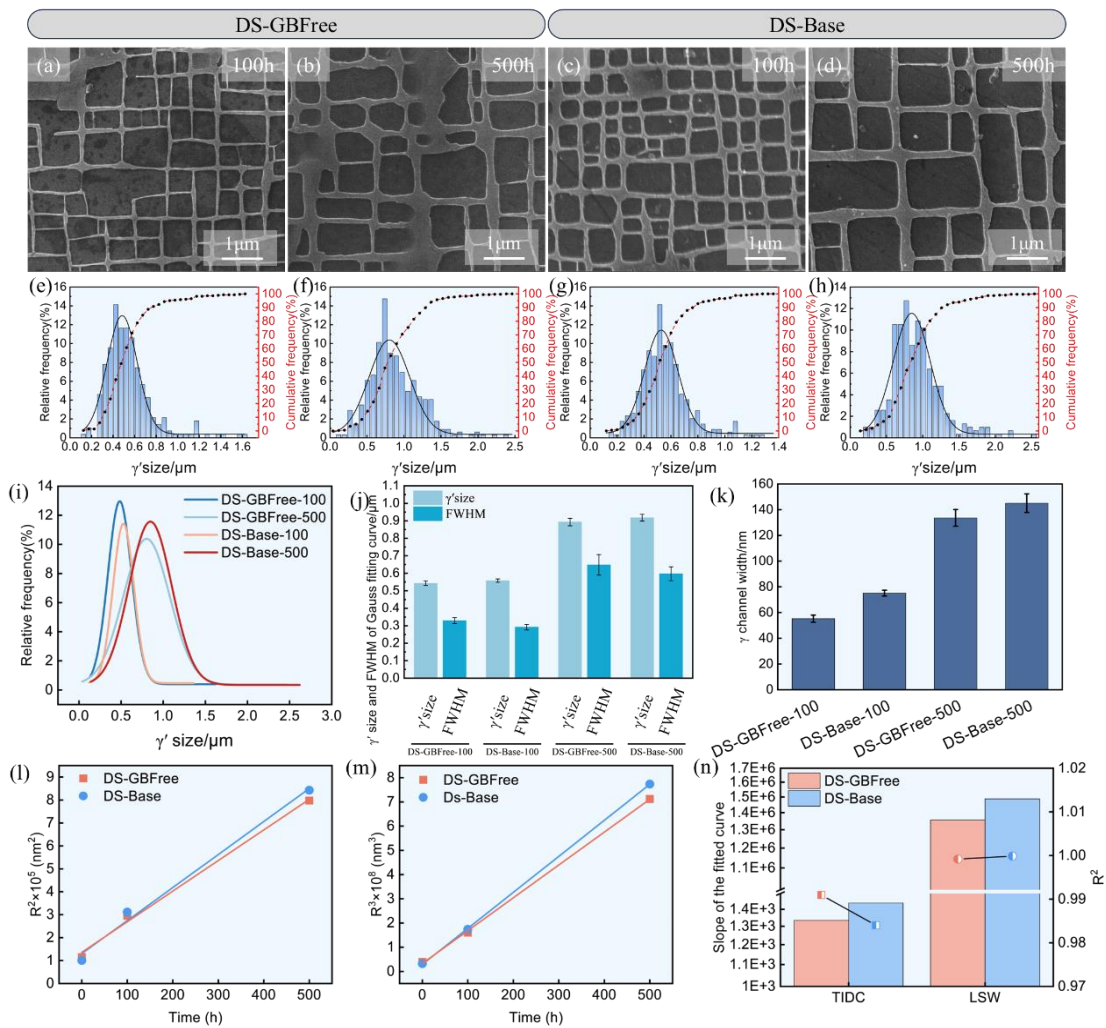

**Figure 4** The γ/γ′ two-phase microstructure after 100h and 500h thermal exposure of (a,b) DS-GBFree and (c,d) DS-Base alloys. The γ′ precipitate size distributions with Gaussian fitting curves of (e,f) DS-GBFree alloy and (g,h) DS-Base alloy. (i) The Gaussian fitting curves, (j) the γ′ average size and FWHM of Gaussian fitting curves and (k) γ-phase channel widths in different regions in DS-GBFree and DS-Base alloy. Plot of (l) $r^2$ and (m) $r^3$ vs. time during aging at 1040°C. (n) The slope and coefficient of determination of the coarsening fitting curve.

Moreover, although DS-Base alloy showed marginally smaller γ′ dimensions (317 ± 3 nm) than DS-GBFree alloy (339 ± 4 nm) in heat-treated condition, this trend was reversed during thermal exposure. The TIDC (Trans-Interface Diffusion-Controlled) and LSW (Lifshitz-Slyozov-Wagner) theories were conventionally employed to describe γ′ phase coarsening behavior in superalloys[36,37], positing that the square (TIDC) or cube (LSW) of γ′ coarsening with aging time. To quantitatively assess post-heat-treatment coarsening in both alloys, linear regression analyses were performed on γ′ size evolution versus time, with results presented in Figure 4(l-n). Figure 4(n) demonstrates superior linear correlation for the cubic relationship ($R^2 > 0.999$) compared to the quadratic fit ($R^2 < 0.99$), confirming LSW model dominance and demonstrating matrix diffusion-controlled coarsening mechanisms. The LSW-derived slopes represent diffusion-controlled constant K, yielding values of $1.36 \times 10^6$ nm³/h for DS-GBFree and $1.49 \times 10^6$ nm³/h for DS-Base. This 8.7% reduction directly evidenced that eliminating grain boundary strengthening elements suppressed matrix diffusion, thereby retarding γ′ phase coarsening.

Comparative analysis of γ-phase channel width evolution revealed similar initial dimensions (~40 nm, <50 nm maximum) in both alloys after heat treatment. However, thermal exposure induced substantial channel widening, with DS-Base alloy exhibiting accelerated coarsening compared to DS-GBFree alloy (Figure 4(k)). After 100h exposure, DS-GBFree alloy demonstrated a moderate 43% increase in γ channel width compared to the dramatic 127% expansion observed in DS-Base alloy. This discrepancy amplified progressively with extended exposure duration, reaching 246% and 338% channel width increments for DS-GBFree and DS-Base alloys respectively after 500h thermal aging. This differential behavior holds particular significance for creep resistance, as early-stage dislocation motion through γ channels strongly correlates with channel dimensions. The driving stress for dislocation movement derives from both applied loads and thermal mismatch stresses[38–40], while the Orowan resistance follows[38]:

$$\tau_{Orowan\ resistance} = \sqrt{\frac{2}{3}\frac{\mu b}{h}} \qquad (1)$$

where μ represents shear modulus, b = 1/2[011] denotes Burgers vector, and h corresponds to γ-channel width. The retarded channel coarsening in DS-GBFree alloy consequently enhances resistance to dislocation glide during creep deformation, demonstrating how grain boundary strengthening element removal modifies microstructural evolution pathways to improve high-temperature performance.

Beyond the differences in γ′ coarsening behavior, thermal exposure induced the formation of $M_6C$ carbides specifically at grain boundaries in DS-Base alloy which was converted from MC carbides, as illustrated in Figure 5. These secondary carbides exhibited distinct elemental preferences compared to primary MC carbides, while MC carbides predominantly contain Ti and Ta (Figure 3), $M_6C$ carbides showed strong enrichment in Mo and W (Figure 5(b-c)). The formation mechanism followed the phase transformation:

$$MC + \gamma \rightarrow M_6C + \gamma' \qquad (2)$$

This reaction created γ-depletion zones surrounding $M_6C$ carbides, with residual MC

particles persisting at carbide interfaces. It is generally recognized that $M_6C$ carbides predominantly form in alloys with elevated Mo and W contents. Their impact on alloy performance critically depended on both spatial distribution and morphological characteristics[41,42]. As demonstrated in Figure 5, coarse and interconnected $M_6C$ carbides typically serve as stress concentration sites during service conditions[41]. In this study, $M_6C$ carbides were exclusively observed at grain boundaries, with no intragranular precipitation detected. This exclusive intergranular precipitation likely stemmed from the combined effects of (1) elevated defect density and (2) accelerated atomic diffusion at grain boundaries. More importantly, this observation confirmed that the strategic removal of grain boundary strengthening elements in DS-Base alloy effectively suppresses the formation of detrimental $M_6C$ carbides along boundaries, thereby mitigating stress concentration risks during service.

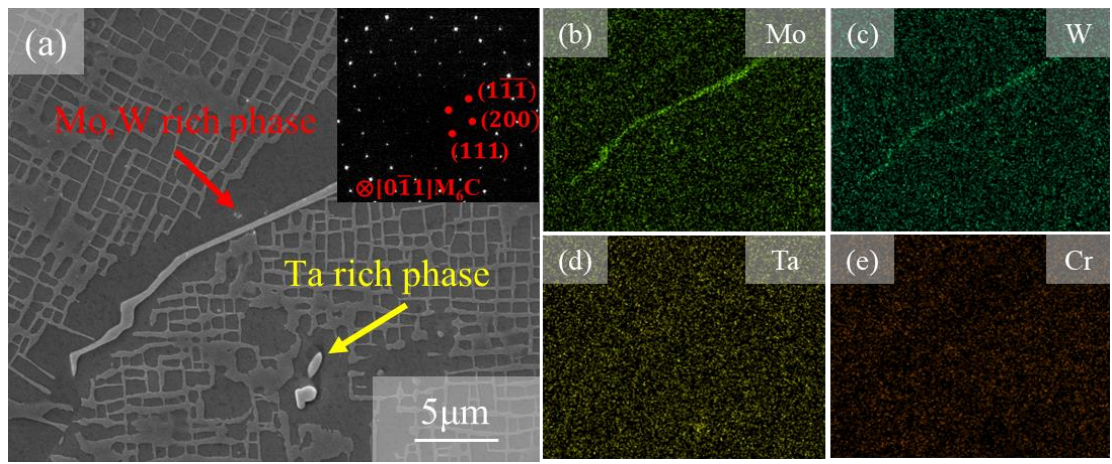

**Figure 5** (a) The morphology of $M_6C$ carbides with SAED (Selected Area Electron Diffraction) patterns precipitated at grain boundary of DS-Base alloy after 200h thermal exposure and (b-e) corresponding EDS elemental mapping analysis.

## 3.3 Creep properties and fracture mechanisms

Figure 6 presented comparative creep curves, strain rate-time relationships, and strain rate-strain profiles for DS-GBFree and DS-Base alloys under 1040°C/137MPa conditions. The rupture lives of DS-Base and DS- GBFree alloys reached 249 h and 403 h, respectively. Both alloys exhibited characteristic three-stage creep behavior: initial strain accumulation with rafting structure formation and dislocation network development, followed by a steady-state stage with minimal strain rate variation ($\approx 3\times 10^{-8}$ s$^{-1}$), and final accelerated deformation leading to fracture. Figure 6(b) reveals comparable minimum creep rates between the alloys, indicating similar strengthening effectiveness. The lifespan discrepancy primarily originated from extended steady-state duration in DS-Base alloy. Notably, grain boundary modification delayed tertiary creep initiation, and the critical time threshold for accelerated creep increases from 200h in DS-Base alloy to 300h in DS-GBFree alloy.

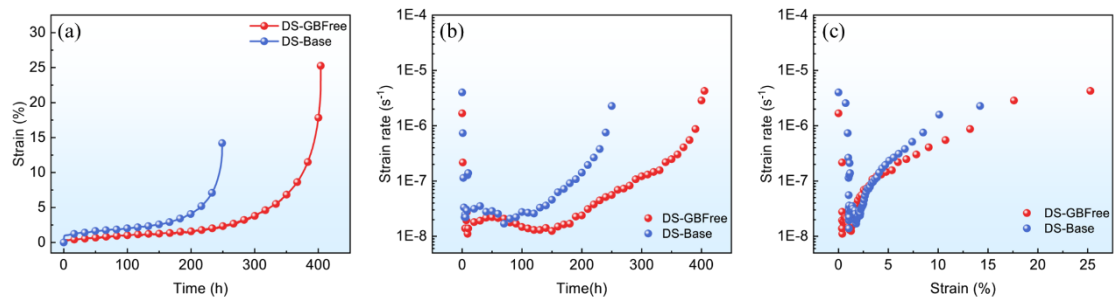

**Figure 6** Comparative analysis of creep curves between DS-GBFree and DS-Base alloys: (a) Strain-time curves; (b) Strain rate-time curves; (c) Strain rate-strain curves.

### 3.3.1. Intragranular fracture mechanisms

Generally, during the steady-state creep stage of nickel-based directionally solidified superalloys under high-temperature/low-stress conditions, the initially cuboidal intragranular γ′ precipitates underwent directional coarsening to form rafted structures, while γ/γ′ interfaces developed dislocation networks[43–46]. Upon entering tertiary creep, massive superdislocation cutting through γ′ precipitates induced localized stress concentration leading to fracture[47]. The raft structure of DS-GBFree and DS-Base alloys was listed in Figure S2. To analyze the micro defects and fracture modes of two alloys after fracture, Figure 7 compared cross-sectional TEM observations near (2 mm) and far from (10 mm) fracture surfaces in both alloys. The superdislocation density in γ′ phase showed significant stress gradient dependence. Notably, DS-Base alloy exhibited substantially lower superdislocation density than DS-GBFree alloy at equivalent positions, despite similar stress concentration levels. This reduced dislocation density likely contributes to its shorter accelerate-state creep duration, suggesting that the removal of grain boundary strengthening elements enhances intragranular γ′ phase resistance to superdislocation penetration.

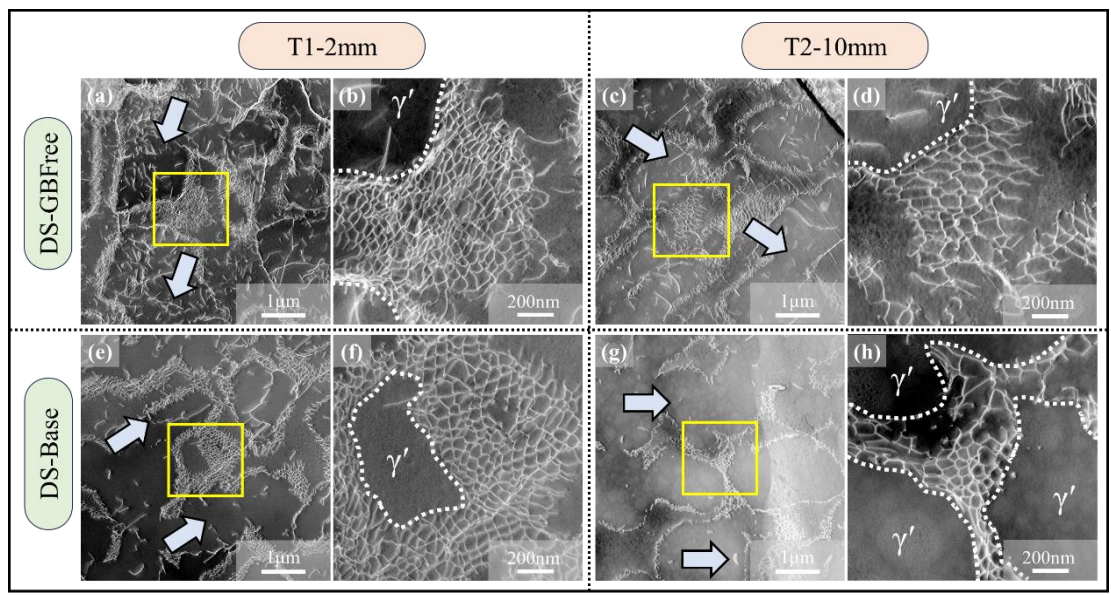

**Figure 7** Dislocation network configurations in both alloys after creep rupture: (a-d) DS-GBFree alloy. (e-h) DS-Base alloy.

Figure 7 also presented magnified views (yellow boxes) of dislocation network configurations at different locations in the two alloys. The networks in regions 10 mm from fracture surfaces demonstrated regular hexagonal or octagonal configurations in both alloys (Figure 7(d)(h)). Near fracture surfaces (2 mm), DS-Base alloy exhibited rhombic network morphologies (Figure 7(f)), while DS-GBFree alloy predominantly showed transitional configurations (Figure 7(b)). Interestingly, Pollock et al.[48] proposed that transition dislocation networks were mostly composed of slip dislocations and therefore exhibit a slightly curved feature. The evolution from transitional to equilibrium configurations involved dislocation rearrangement through two concurrent processes: (1) dislocation reaction-driven line energy minimization and (2) misfit stress-driven configuration stabilization. Equilibrium networks typically displayed straight dislocations arranged in low-energy configurations (octagons, hexagons, or rhombuses). However, in many studies[49,50], transitional dislocation networks also existed in samples during the steady-state creep stage or after creep rupture. The persistence of transitional networks in post-creep specimens, suggested incomplete equilibrium attainment. This phenomenon likely aroused from localized interfacial misfit disequilibrium during steady-state creep, where dynamic competition between dislocation generation and annihilation maintains metastable network configurations.

Statistical analysis of γ channel widths in longitudinal sections after creep rupture (Figure 8(a)) revealed critical insights into microstructural evolution during deformation. The conical representations (blue for DS-GBFree and red for DS-Base) illustrated width variations with distance from fracture surfaces. Both alloys develop widened γ channels near high-stress fracture regions, while maintaining relatively stable dimensions in areas farther from the rupture. Notably, DS-GBFree alloy consistently displayed larger channel widths than DS-Base alloy across all positions, a finding that appears contradictory to the thermal exposure results in Figure 4.

This discrepancy was resolved through a 250h creep interruption test on DS-GBFree alloy, considering that the creep life of the two alloys differed by more than 100h. The creep interruption test protocol involved terminating the experiment after 250 hours of continuous loading while maintaining the applied stress throughout the cooling process until the specimen reached ambient temperature. Since the interrupted test was short in time, no necking and stress concentration occurred, so the structures in different parts of the specimen were basically consistent. The γ channel widths of interruption test became uniform and were positioned below all conical peaks. This confirmed that under identical loading timeframes, DS-Base alloy developed wider γ channels than DS-GBFree alloy. The accelerated channel widening in DS-Base correlated with its significantly shorter creep life (>150 hours less than DS-GBFree alloy). Mechanistically, the broader γ channels in DS-Base facilitate enhanced dislocation motion and propagation, creating preferential pathways for plastic deformation that ultimately lead to premature creep fracture.

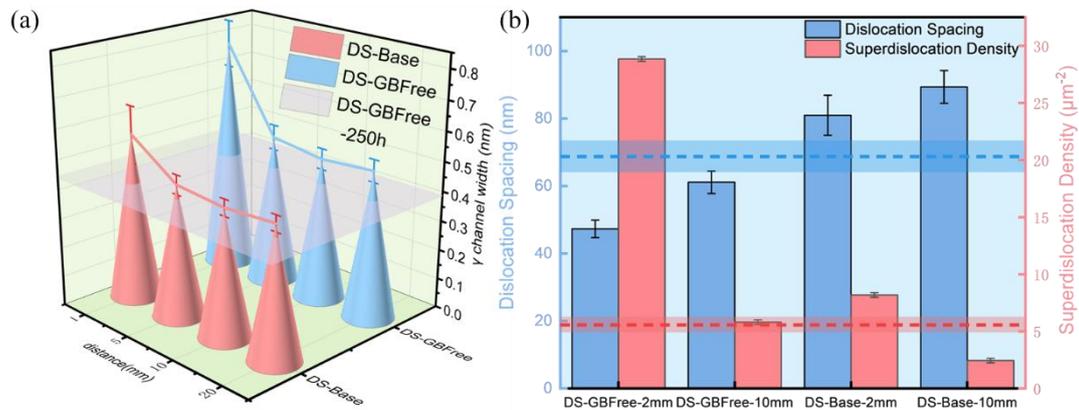

**Figure 8** Quantitative characterization of microstructural parameters in both alloys after creep rupture at different locations relative to fracture surface: (a) γ-channel width measurements; (b) Superdislocation density and dislocation network width with blue and red dash lines representing the dislocation spacing and superdislocation density of samples of DS-GBFree alloy interrupted after 250h creep with error range.

To comparatively investigate the creep fracture mechanisms of both alloys, quantitative analysis of dislocation network spacing and superdislocation density was conducted using data from Figure 7, with statistical results presented in Figure 8(b). For temporal-equivalent comparison, microstructural parameters from the 250h-interrupted DS-GBFree alloy specimen (represented by blue and red dashed lines in Figure 8(b), accompanied by light-colored bands indicating measurement error) were integrated with rupture data. The corresponding dislocation structures were illustrated in Figure S4. Both alloys exhibited progressive refinement of dislocation networks with proximity to fracture surfaces, a phenomenon attributed to stress concentration effects near rupture[62]. The temporal evolution of microstructure revealed that extended creep duration promoted dislocation network densification[47,51], as evidenced by the larger network spacing in 250h-interrupted DS-GBFree alloy specimens compared to their ruptured specimens, though still smaller than those observed in DS-Base alloy. Furthermore, the established correlation between γ/γ′ lattice mismatch and dislocation network density during steady-state creep was quantified through the following constitutive relationship[63]:

$$|\delta_{eff}| = \frac{|b|}{d_{dn}} \quad (3)$$

where $\delta_{eff}$ represented the effective lattice misfit, $d_{dn}$ was the spacing of the dislocation network. The Burgers vector was taken as $a/2[\bar{1}01]$. The systematically reduced dislocation network spacing observed in DS-GBFree alloy across various microstructural locations compared to DS-Base alloy resulted in greater absolute values of lattice mismatch. This enhanced mismatch magnitude typically corresponded to elevated coherency stresses at γ/γ′ interfaces. The formation of dense dislocation networks effectively alleviated these interfacial coherency stresses through localized strain accommodation. The experimental correlation between network density and lattice mismatch provided a micromechanical basis for understanding the >100h creep life advantage of DS-GBFree alloy over DS-Base alloy.

During tertiary creep deformation, dislocation networks progressively lost their stress accommodation capacity, undergoing progressive destabilization that triggered topological inversion of the γ/γ′ microstructure[53,54]. This structural evolution manifested in post-rupture cross-sections as fragmented γ′ precipitates adopting discrete island morphologies, accompanied by extensive superdislocation sheared penetration into the γ′ phase. To quantitatively assess this microstructural degradation, superdislocation density can be determined through[55]:

$$\lambda = \frac{N}{S} \qquad (4)$$

where $\lambda$ represented the superdislocation density, N was the terminal point number of dislocations shearing into the γ′ precipitate, and S represented the area of the statistical region. The statistical results of super dislocation density under different conditions were shown in Figure 8(b). The observed superdislocation density paradox in DS-GBFree alloy challenged conventional understanding of creep mechanisms in superalloys. Generally, for single crystal superalloys, the alloys with longer life had fewer superdislocations after fracture because the dislocation network was a stronger barrier to superdislocation shearing[53,54]. For both samples interrupted after 250h and those after creep rupture, the superdislocation density in the DS-GBFree alloy consistently exceeded that in the DS-Base alloy. This demonstrates enhanced resistance of the γ′ phase to superdislocations in the DS-GBFree alloy, a conclusion previously reported in our research group's earlier publication[56].

### 3.3.2. Grain boundary fracture mechanisms

Contrary to conventional understanding that grain boundary strengthening elements (C, B, Zr) enhanced alloy performance, this study demonstrated improved properties through their strategic removal. To unravel the underlying mechanisms dictating this performance inversion and to characterize the strength relationship between grain boundaries and intracrystalline, EBSD analysis was conducted on post-fracture microstructures at different distances from fracture surfaces in both alloys. Figure 9 presented Kernel Average Misorientation (KAM) maps and Inverse Pole Figure (IPF) maps under identical stress orientation (indicated by light yellow arrows).

The KAM maps revealed significant stress concentration and strain localization near fracture surfaces in both alloys. The DS-Base alloy demonstrated non-selective stress concentration distribution, exhibiting significant accumulation at both grain boundaries and within grains. This indicated extensive homogeneous plastic deformation within the γ/γ′ microstructure prior to rupture. In stark contrast, Following the removal of boundary-strengthening elements, the DS-GBFree alloy displayed position-dependent stress concentration characteristics as shown in Figure 9 (a), with pronounced accumulation at grain boundaries compared to relatively lower stress levels within grains. This pattern signifies that plastic deformation and damage accumulation were largely confined to the GB regions.

Notably, both alloys developed similar microcrack morphologies along γ/γ′ interfaces (Figure 9 (c)(d)(n)). In DS-Base alloy, microcracks predominantly initiated at γ/γ′ interfaces rather than carbide/matrix interfaces, even near carbide particles (Figure 9(n)). This contrasted with conventional observations where carbides typically induced localized stress concentrations by restricting grain boundary migration and hindering dislocation motion[57–59]. And this observation confirmed that neither grain boundary nor intragranular carbides act as stress concentration sites in DS-Base alloy. This indicated that part of the reason for the shorter creep life of DS-base alloys was due to the lower intragranular strength compared to DS-GBFree.

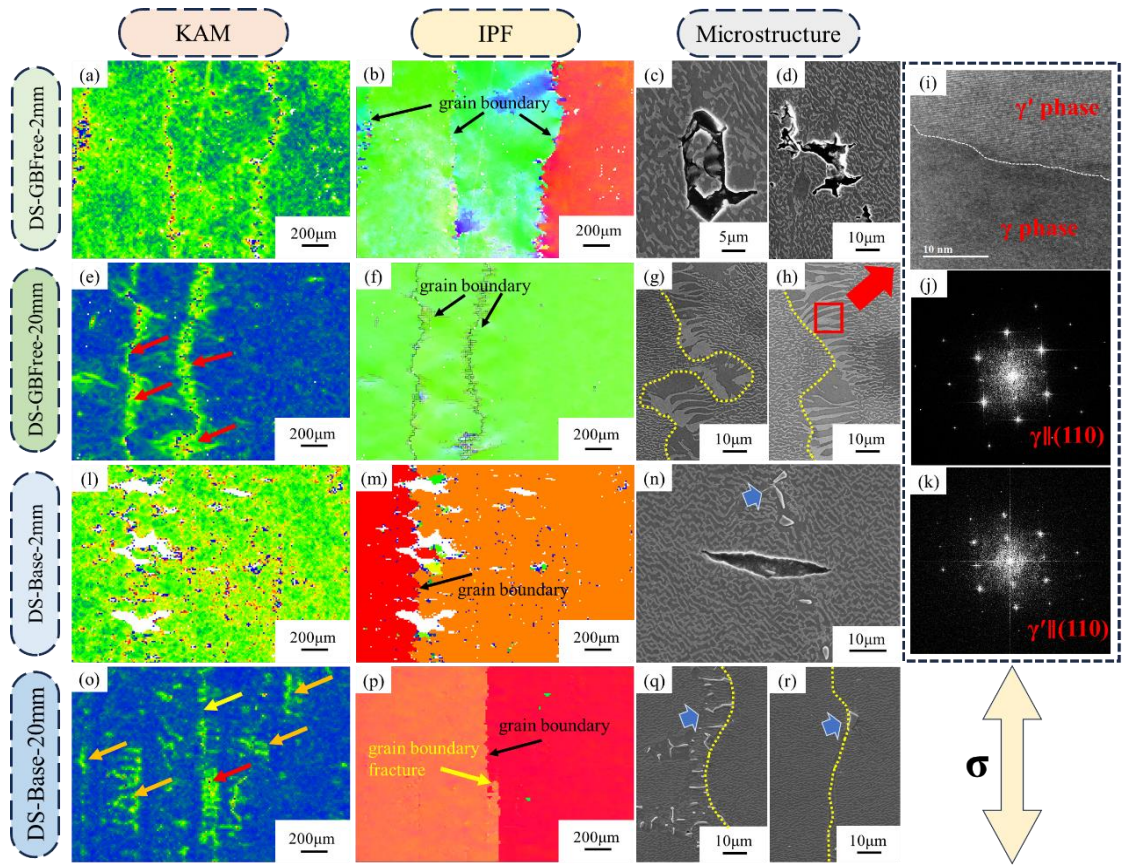

**Figure 9** Kernel Average Misorientation (KAM), Inverse Pole Figure (IPF) mapping analyses and microstructure of both alloys after creep rupture: (a-h) DS-GBFree alloy; (l-r) DS-Base alloy. (i-k) HRTEM and FFT image of discontinuous precipitation on grain boundary.

In order to further explore the differences in fracture mechanisms between the two alloys, the microstructure at the far fracture site of the two alloys were analyzed. Both alloys exhibited similar stress concentration patterns in regions distal to fracture surfaces compared to near-fracture areas. As shown in Figure 9(e), the DS-GBFree alloy demonstrates significant stress concentration at grain boundaries despite the absence of grain boundary strengthening elements and carbide precipitates that typically impede boundary motion. This phenomenon may originate from discontinuous precipitation at grain boundaries during creep, inducing grain boundary migration and subsequent formation of serrated grain boundaries (Figure 9(g)(h)) that effectively inhibit intergranular sliding, thereby enhancing creep resistance. The discontinuous precipitation at grain boundaries was identified as γ and γ′ phase, as shown in Figure 9(i-k). These serrated boundaries in post-creep DS-GBFree alloy contrast markedly with the straight boundaries observed in heat-treated conditions.

In DS-Base alloy, localized stress concentration occurred not only at intragranular regions (orange arrows in Figure 9(o)) but also at specific grain boundaries (red arrows in Figure 9(o)), while strain-free zones coexisted at adjacent boundaries (yellow arrows). Different from DS-GBFree alloy, intragranular regions displayed stress levels comparable to those at boundaries, and stress accumulation at distal grain boundaries in DS-Base results not from carbide presence but rather from mechanical interlocking between neighboring grains, as shown in Figure 9(p). The distinct grain boundary morphology in DS-Base (Figure 9(q)(r)) reveals that carbide pinning effectively suppresses both boundary migration and discontinuous precipitation. While this pinning mechanism stabilizes grain boundaries, it simultaneously inhibits the formation of performance-enhancing serrated boundaries observed in DS-GBFree alloy. The above results showed that the DS-base alloy not only exhibited lower intragranular strength than the DS-GBFree alloy, but also exhibited lower grain boundary strength due to the absence of serrated grain boundaries.

### 3.3.3. Grain boundary serration

To quantitatively characterize grain boundary curvature in both alloys, fast Fourier transform (FFT) analysis was systematically conducted as shown in Figure 10, with white arrows indicating the directional solidification direction and creep stress orientation. Significant microstructural differences were observed in the heat-treated condition (Figures 10(a)(d)). The DS-Base alloy exhibited naturally grain boundaries serration likely formed during solidification, decorated with blocky MC-type carbides (yellow arrows). In contrast, the removal of boundary-strengthening elements in the DS-GBFree alloy suppressed curvature formation during solidification, resulting in straight boundaries aligned with the directional solidification direction. FFT analysis (Figure 10(g)) revealed distinct amplitude-wavelength distributions between the two alloys. The DS-Base alloy demonstrated consistently higher amplitudes across all wavelengths compared to the DS-GBFree alloy. As shown in the distribution curves (right panel of Figure 10(g)), DS-Base exhibits relatively uniform amplitude distribution with an average value of approximately 5 μm, while the DS-GBFree alloy displays significantly lower amplitudes primarily concentrated around 1 μm.

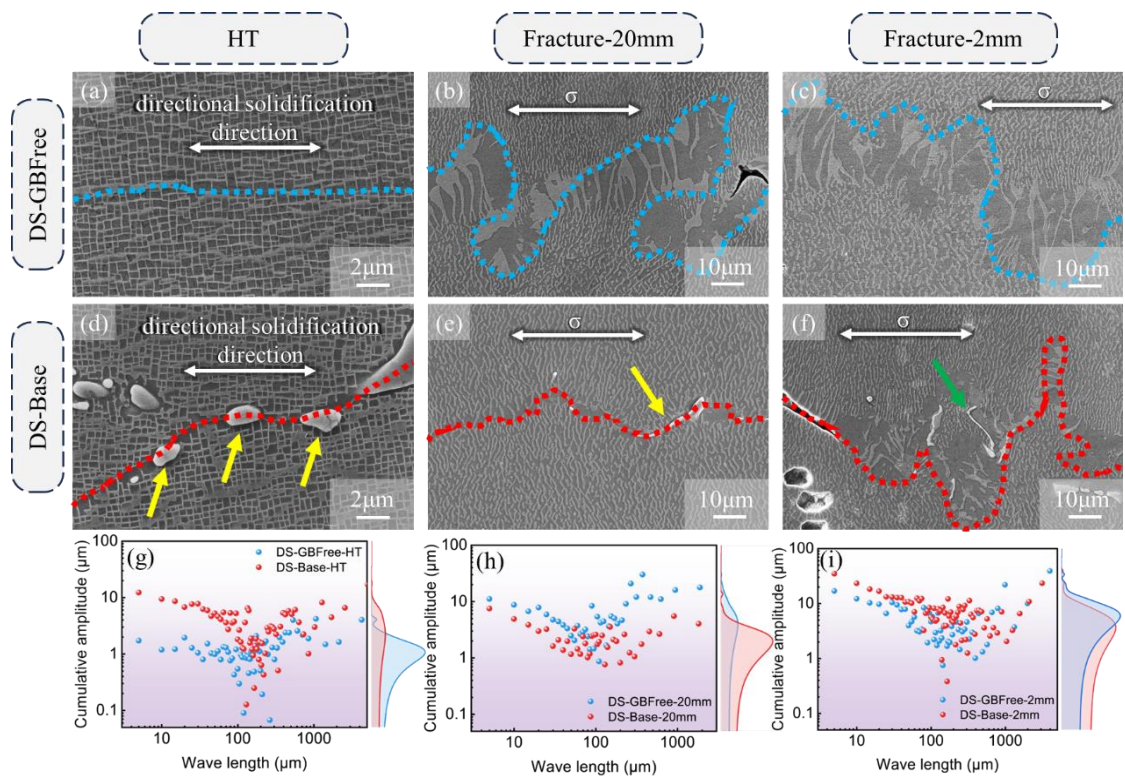

**Figure 10** Grain boundary characteristics and Fourier transform analysis of DS-GBFree and DS-Base alloys under various conditions: (a-c) DS-GBFree alloy microstructures in heat-treated state, 20 mm and 2 mm regions from creep fracture surface; (d-f) DS-Base alloy microstructures in heat-treated state, 20 mm and 2 mm regions from creep fracture surface; (g-i) Fast Fourier Transform (FFT) calculation results of grain

boundaries of two alloys in heat-treated state, 20 mm and 2 mm post-creep regions with the amplitude distribution curve on the right of each picture.

Figure 10 (b) and (e) presented grain boundary morphologies in regions 20 mm from fracture, where no significant stress concentration occurred. Following prolonged creep exposure, the DS-Base alloy maintained serrated boundaries with amplitude-wavelength profiles similar to its heat-treated condition. In contrast, the DS-GBFree alloy developed pronounced grain boundaries serration (indicated by blue dashed lines in Figure 10(b)) in distant fracture regions after creep loading. The FFT statistical plots (Figure 10(h)) revealed a marked upward shift in amplitude distribution across all wavelengths for post-creep DS-GBFree alloy compared to its heat-treated state. This curvature evolution effectively inhibited grain boundary migration, significantly enhancing creep resistance[60]. The observed boundary curvature correlated directly with stress concentration near grain boundaries shown in Figure 9(e), confirming the critical role of serrated boundary formation in mechanical performance improvement.

Grain boundary morphologies in high-stress concentration regions 2 mm from fracture surfaces were presented in Figure 10(c) and (f). In DS-Base alloy, compared to non-stressed regions distant from the fracture, stress concentration likely promoted grain boundary migration, forming serrated grain boundaries indicated by red curves. $M_6C$ carbides adjacent to these serrated boundaries were observed as shown by green arrows in Figure 10(f). In contrast, stress concentration showed negligible influence on grain boundary morphology in DS-GBFree alloy, maintaining configurations similar to those in distant fracture regions (Figure 10(b)). FFT statistical results in Figure 10(i) demonstrated differential effects of stress concentration on grain boundary serration between the alloys. DS-GBFree alloy exhibited minor amplitude increase across various wavelengths under high stress concentration, while DS-Base displayed significant curvature enhancement. The stressed DS-Base boundaries show substantially increased amplitudes compared to non-stressed regions, upward-shifted distribution curves and amplitude-wavelength characteristics exceeding those of DS-GBFree alloy.

## 4. Discussion

*4.1 Effect of removal of grain boundary strengthening elements on grain boundary strength*

In this study, the strategic removal of grain boundary strengthening elements (C, B, Zr) prevented carbide precipitation particularly at grain boundaries in the DS-Base alloy (Figure 3(e)), while also suppressing carbide phase transformations during thermal exposure (Figure 5(a)). High-temperature creep testing revealed that eliminating these grain boundary strengthening elements significantly enhanced the alloy's creep life, as demonstrated in Figure 6. Microstructural characterization of grain boundaries showed that although precipitates pinned grain boundaries in the DS-Base alloy (Figure 9(n)(q)(r)), cracks neither initiated nor propagated around these carbides. Furthermore, no significant stress concentration occurred near grain boundaries compared to grain interiors, as evidenced in Figure 9(l)(n). Conversely, the DS-GBFree alloy fabricated without grain boundary strengthening elements developed large-amplitude serrated grain boundaries (Figure 9(g)(h)) that strengthened these interfaces. This configuration facilitated pronounced stress concentration adjacent to grain boundaries.

Quantitative analysis of serrated grain boundary revealed that although DS-Base alloy possessed low-amplitude serrated grain boundaries in the heat-treated condition, these boundaries do not propagate during service. Significant curvature development only occurs in stress-concentrated regions during the final creep stage. The original low-amplitude serrated boundaries failed to impede dislocation propagation and boundary sliding, as evidenced by the absence of stress concentration at grain boundaries in Figure 9(o) after creep fracture, ultimately leading to transgranular failure (longitudinal section near fracture surface shown in Figure S3). In contrast, the removal of boundary-strengthening elements in DS-GBFree alloy eliminated initial curvature in the heat-treated state but enabled non-selective grain boundary serration formation during service. These stress-induced curved boundaries effectively inhibited boundary migration while enhancing grain boundary strength, resulting in intergranular fracture (longitudinal section near fracture surface shown in Figure S3). This demonstrated that eliminating boundary-strengthening elements not only maintained but improved grain boundary strength through promoted curvature formation.

## 4.2 Effect of removal of grain boundary strengthening elements on intragranular strength

### 4.2.1 Effect on heat-treated phase structure

The creep resistance of superalloys depended not only on grain boundary strength but also on intragranular strengthening in DS superalloys[61]. The results in Figure 9 showed that cracks in the DS-base alloy initiated preferentially from the intragranular $\gamma/\gamma'$ phase interface. And the $\gamma'$ phase content directly correlated with creep performance[62,63], as variations in $\gamma'$ content significantly altered its morphological characteristics. Figures 1 and 2 demonstrated that the removal of grain boundary strengthening elements substantially affected both the size and area fraction of $\gamma'$ precipitates, despite these elements not being $\gamma'$-forming elements. The influence of carbon on dendritic segregation remained controversial in previous research[24,33]. In this study, both alloys exhibited comparable two-phase microstructures and segregation coefficients after heat treatment. Notably, DS-GBFree alloy demonstrated slightly greater segregation intensity than DS-Base alloy, attributable to the elimination of grain boundary strengthening elements in DS-GBFree alloy which liberated Ta and Ti, principal constituents of MC carbides, thereby increasing the segregation coefficients, particularly for Ta. The observed increase in segregation coefficients for Al and Co after heat treatment arises from differences in composition analysis: cast state measurements employed selective point analysis that deliberately excluded eutectic constituents, with interdendritic sampling positioned adjacent to eutectic regions, thereby underestimating Al content. For Co element, solidification-induced rejection during eutectic formation enriched surrounding regions, creating composition profiles resembling dendrite cores and consequently depressing measured segregation coefficients[32].

To further investigate this unexpected influence on γ′ content and thermodynamic properties, equilibrium phase diagrams of DS-GBFree and DS-Base alloys were calculated using Thermo-Calc software, as shown in Figures 11(a) and 10(b) respectively. Notably, MC carbides exhibit thermal instability at elevated temperatures and consequently decompose, leaving only $M_6C$ carbides in the phase diagram of DS-Base alloy. This phase evolution was consistent with the results of the thermal exposure test of the DS-Base alloy. Furthermore, grain boundary strengthening elements would form low melting point phases and were therefore removed in some second-generation single crystals[64,65]. In this study, the removal of grain boundary strengthening elements significantly elevated the alloy's solidus temperature from 1300°C to 1358°C, as evidenced by the initial temperature difference of purple dash-dot lines in Figure 11. This modification created an expanded solutioning window for DS-GBFree alloy. Although identical maximum solutioning temperatures were intentionally set in this study to compare segregation effects, the enhanced thermal stability enabled potential increases in DS-GBFree alloy's solutioning temperature, indicating greater performance optimization potential.

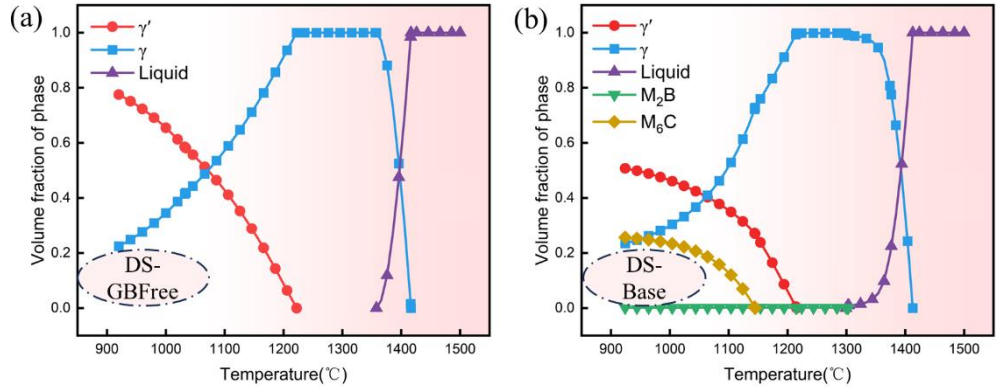

**Figure 11** Thermodynamic phase diagram calculations of (a) DS-GBFree alloy and (b) DS-Base alloy. (The data of DS-GBFree was adapted and re-plotted from Ref.[32]).

In addition, the removal of grain boundary strengthening elements not only removed the carbide phase and boride phase in the DS-GBFree alloy phase diagram, but also greatly increased the content of γ′ phase at different temperatures, although it did not change the temperature at which the γ′ phase was completely dissolved, which is consistent with the conclusion in Figure 1. Above all, the elimination of grain boundary strengthening elements achieved three critical modifications: 1) Complete removal of carbide and boride phases from the DS-GBFree alloy's phase diagram; 2) The significant expansion of the solution window, which provided space for increasing the alloy solution temperature and improving the degree of homogenization; 3) Significant enhancement of γ′ phase content across various temperatures; 4) Preservation of the γ′ solvus temperature. This synergistic effect demonstrated that removing boundary-strengthening elements effectively increased volume fraction of γ′ phase, thereby enhancing the alloy's intragranular strength and consequently improving its creep resistance through matrix strengthening mechanisms.

### 4.2.2 Effect on intragranular creep resistance

Figure 2 and Figure 4 revealed an inversion in γ channel width between the two alloys under different conditions: DS-GBFree alloy exhibited wider γ channels in the heat-treated state, while DS-Base alloy showed broader channels after thermal exposure. This reversal indicated that removing grain boundary strengthening elements enhanced γ channel width stability. This strengthening effect was clearly demonstrated in Figure 8. After equivalent loading durations, the DS-GBFree alloy exhibited narrower γ channels, providing greater resistance to dislocation motion within these pathways. Furthermore, prolonged thermal exposure of the DS-GBFree alloy led to the formation of $M_6C$-type carbides at grain boundaries (Figure 3), which absorbed strengthening elements such as Mo and W from the γ matrix. This elemental depletion reduced the lattice misfit between γ and γ' phases, thereby diminishing intragranular strength.

Besides, the grain boundary serration in DS-GBFree alloy fundamentally altered damage progression during tertiary creep, extending the acceleration stage duration to 100h compared to 50h in DS-Base alloy (Figure 6), resulting in the much higher superdislocation density of DS-GBFree alloy after fracture than that of DS-Base alloy. This prolonged deformation phase enabled progressive accumulation of superdislocations through boundary-participated strain localization, as certified by EBSD analysis showing heterogeneous strain distribution predominated in DS-GBFree alloy, in which the strain mainly occurred at the grain boundaries, versus homogeneous distribution with weak grain boundary in DS-Base (Figure 9(a)(l)). This dual mechanism of coordinated deformation within the grain and at the grain boundary explained DS-GBFree alloy's exceptional creep life.

## 4.3 Correlation between grain boundary strengthening elements and the microstructure and properties

As shown in Figure 12(a), both Re-free 1st DS superalloys demonstrate rupture lives comparable to 2nd SC superalloy René N5 (≈300 h). Furthermore, since both alloys lacked Re, their raw material costs, even when considering only this factor, remained comparable to the first-generation directionally solidified alloy MGA1400, yet substantially lower than second-generation single-crystal superalloys, as illustrated in Figure 12(b). The DS-GBFree alloy demonstrated superior creep resistance not only in Larson-Miller (LM) parameter analysis but also under specific service conditions when benchmarked against René N5 superalloy[56,66]. The enhanced creep resistance in DS-GBFree alloy, surpassing René N5 performance, highlighted the effectiveness of strategic grain boundary engineering through elemental removal. This performance breakthrough challenges conventional wisdom that DS superalloys typically lag SC superalloy counterparts by one generation in high-temperature capabilities.

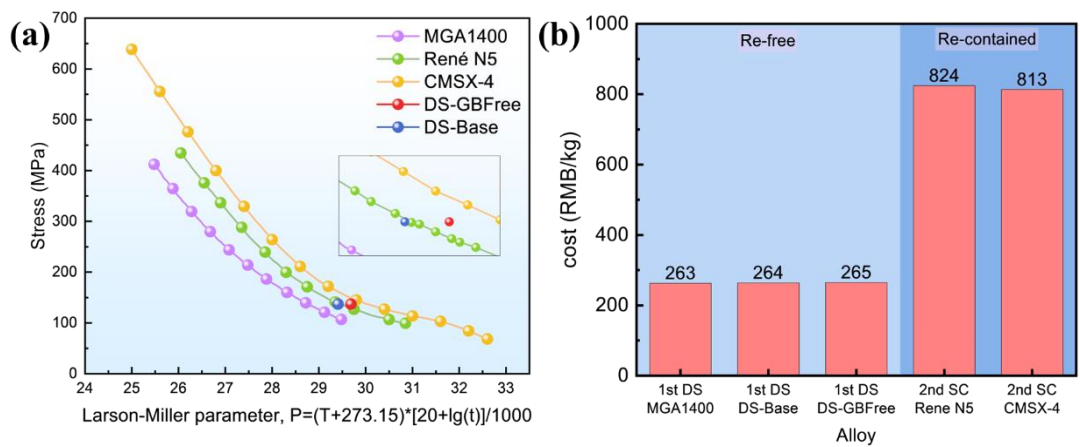

**Figure 12** (a) Creep performance comparison between 1$^{st}$ DS superalloys (DS-GBFree and DS-Base) and two 2$^{nd}$ SC superalloys[47,64,65]. (The creep data of DS-GBFree was adapted and re-plotted from Ref.[56]). (b) Raw material cost estimation for first-generation directionally solidified alloys (Re-free) and second-generation single-crystal superalloys (Re-containing) in (a).

Figure 13 schematically illustrated the influence of grain boundary strengthening elements on microstructural evolution and deformation mechanisms during creep in this study. The 1$^{st}$ generation DS superalloy DS-Base, containing conventional grain boundary strengthening elements, exhibited curved grain boundaries in the heat-treated condition but demonstrated reduced γ′ phase content (69 vol% vs. 75 vol% in DS-GBFree alloy). Conversely, the innovative 1st generation DS superalloy DS-GBFree eliminated the grain boundary strengthening elements, displayed straight grain boundaries and enhanced γ′ phase precipitation in the heat-treated state. Microstructural stability analysis revealed that DS-GBFree alloy maintained superior resistance to γ channel widening during thermal exposure by inhibiting the diffusion and reducing diffusion-controlled constant K of LSW model, showing only 43% increase in channel width compared to 127% for DS-Base alloy of 100h exposure.

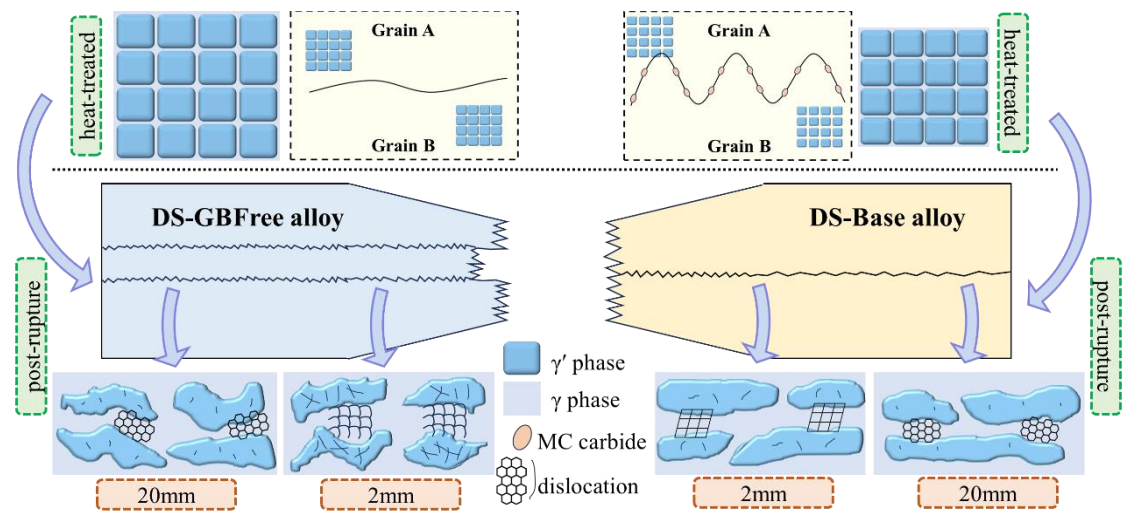

**Figure 13** Schematic illustration of interrelationships between grain boundary strengthening elements and microstructural evolution with underlying deformation mechanisms.

While both alloys exhibited creep properties comparable to second-generation single-crystal superalloys, their deformation mechanisms diverged significantly. DS-Base alloy developed localized high-amplitude curved boundaries near fracture zones during creep, accompanied by widened γ channels at all distances from fracture. DS-GBFree alloy formed position-independent serrated grain boundaries that provide superior boundary strengthening compared to carbide phases. The combination of enhanced lattice misfit and thermal stability enables two critical effects: 1) Formation of finer γ/γ′ interfacial dislocation networks leading to the effective obstruction of superdislocation shear into γ′ precipitates. 2) The coordinated deformation of grain boundaries and intragrain led to more super dislocations in the γ′ phase. These mechanisms synergistically extend the steady-state creep duration by 50% compared to DS-Base alloy while promoting coordinated grain boundary/intragranular deformation that prolongs tertiary creep stage from 50h to 100h. The results demonstrate that strategic removal of grain boundary strengthening elements can paradoxically enhance creep resistance through combined interface strengthening and strain redistribution. Therefore, we proposed that in the future development of Ni-based DS superalloys, the strategy removal of grain boundary strengthening elements may be considered to achieve potentially enhanced creep life.

## 5. Conclusions

This study presented the innovative development of two first-generation directionally solidified (DS) superalloys through strategic elimination of conventional grain boundary strengthening elements, achieving simultaneous enhancement of microstructural stability and creep resistance. A comprehensive characterization protocol was implemented to investigate: (1) initial microstructural configurations in both as-cast and heat-treated conditions, (2) phase evolution during thermal exposure, (3) boundary/intragranular deformation mechanisms during creep, and (4) quantitative assessment of grain boundary curvature evolution across different service stages. The principal findings can be summarized as follows:

(1) The strategic removal of grain boundary strengthening elements liberated Ta and Ti from MC carbides, thereby increasing the eutectic structure area fraction in as-cast alloys from 0.9% to 1.4%. This compositional redistribution enhanced γ′ phase volume fraction (69% vs. 75%). Concurrently, the alloy's solidus temperature increased by 57°C (1300°C → 1357°C), expanding the solutioning window and enabling potential optimized heat treatment protocols.

(2) The strategic removal of grain boundary strengthening elements significantly enhanced microstructural stability during thermal exposure, exhibiting minimal influence on the full width at half maximum (FWHM) of γ′ phase size distribution curves while markedly suppressing γ channel widening and γ′ phase size increasing. Crucially, this compositional adjustment completely inhibited the deleterious phase transformation of MC carbides to $M_6C$ carbides.

(3) The strategic removal of grain boundary strengthening elements, while removing blocky MC carbides at boundaries, paradoxically enhanced boundary strength during creep through two synergistic mechanisms: (a) formation of position-independent serrated grain boundaries, and (b) suppression of boundary sliding. This microstructural reorganization fundamentally altering fracture modes from transgranular to intergranular dominance while maintaining creep resistance.

(4) The strategic removal of grain boundary strengthening elements induced dual-scale intragrain strengthening mechanisms during creep deformation: (1) Suppressed rafting channel widening, effectively impeding dislocation glide within channels; (2) Enhanced γ/γ′ lattice misfit promoted formation of high-density interfacial dislocation networks with refined spacing. This multi-scale strengthening architecture, together with the serrated grain boundary, elevated the critical stress for superdislocation shear of γ′ precipitates, reducing steady-state creep rate, enabling the γ′ phase to accommodate more superdislocations under the grain boundary/intragranular cooperative deformation mechanism, and ultimately prolonged the fracture life.

(5) We advocated that the strategic elimination of grain boundary strengthening elements, which enhanced microstructural stability while enabling synergistic regulation of grain boundary and intragranular strength, should be considered in future development of Ni-based DS superalloys to achieve potential enhancements in creep life.

# Declaration of competing interest

The authors declare that they have no known competing financial interests or personal relationships that could have appeared to influence the work reported in this paper.

# Acknowledgements

This work was jointly supported by the National Key Research and Development Program of China (2024YFB4105000), the National Natural Science Foundation of China (52101155, 52301178), and the Natural Science Foundation of Zhejiang Province (LQ23E010006).

Supplementary materials for

# A Novel Strategy to Strengthen Directionally Solidified Superalloy Through Grain Boundary Simplified Design

Yunpeng Fan[a], Xinbao Zhao[a,b,*], Yu Zhou[a], Quanzhao Yue[a], Wanshun Xia[a,*], Yuefeng Gu[a,b,*], Ze Zhang[a,b]

[a] Institute of Superalloys Science and Technology, School of Materials Science and Engineering, Zhejiang University, Hangzhou 310027, China

[b] State Key Laboratory of Silicon and Advanced Semiconductor Materials, School of Materials Science and Engineering, Zhejiang University, Hangzhou, 310027, China

* Corresponding authors.

E-mail addresses: superalloys@zju.edu.cn (X.B. Zhao), guyf@zju.edu.cn (Y.F. Gu), wanshunxia@zju.edu.cn (W.S Xia)

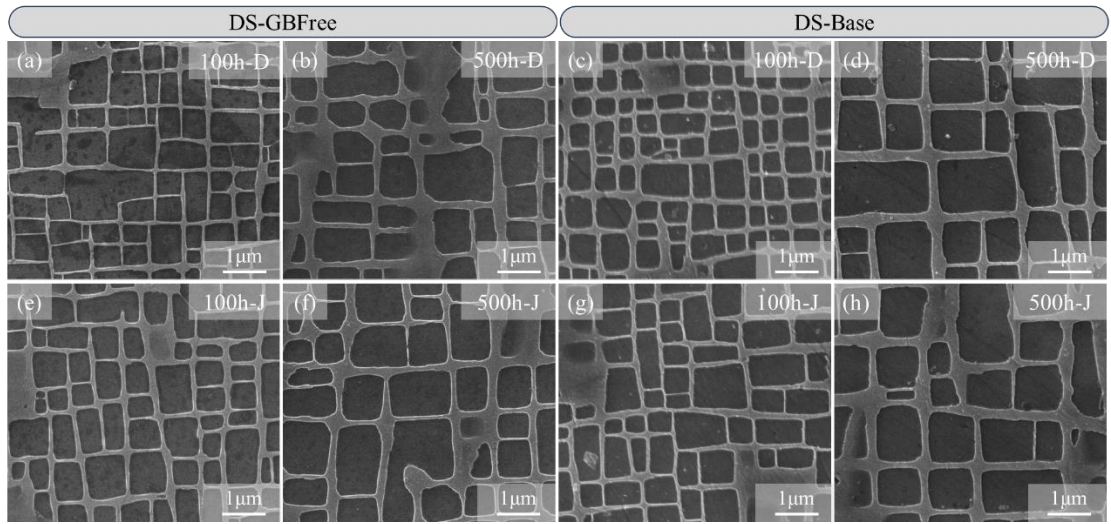

**Figure S1** The γ/γ′ two-phase microstructure after 100h and 500h thermal exposure of (a,b,e,f) DS-GBFree and (c,d,g,h) DS-Base alloys.

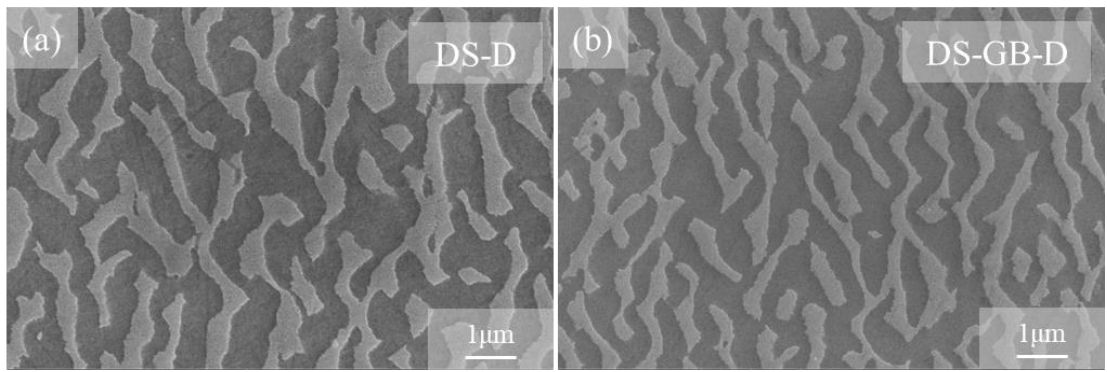

**Figure S2** The raft structure in the dendrite core of the (a) DS alloy and (b) DS-GB alloy.

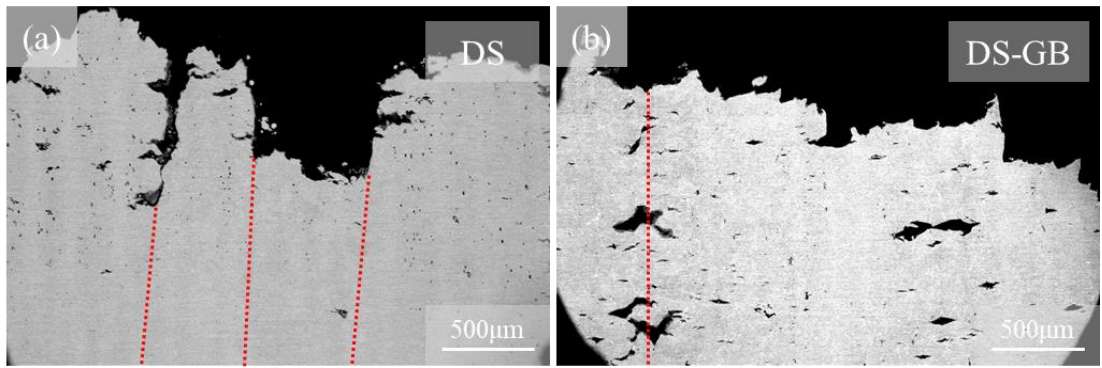

**Figure S3** The longitudinal cross-section morphology of the two alloys near the fracture, with red dash lines indicated grain boundary. (a) DS alloy with intergranular fracture, (b) DS-GB alloy with transgranular fracture.

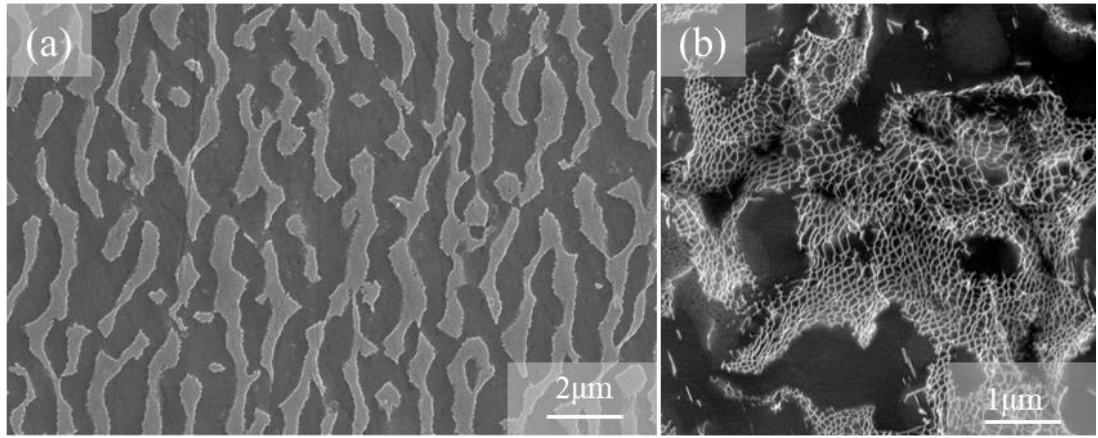

**Figure S4** (a) Rafted microstructure and (b) dislocation network at the interface and super dislocations in the γ′ phase of DS alloy after interruption of creep for 250 h.